\documentclass[aps,preprint,onecolumn,floatfix,superscriptaddress,nofootinbib]{revtex4-1}
\usepackage[T1]{fontenc} % if needed

\newcommand{\sv}{\ensuremath{\langle\sigma v\rangle}}

\newcommand{\mev}{\ensuremath{\,\mathrm{MeV}}}
\newcommand{\gev}{\ensuremath{\,\mathrm{GeV}}}
\newcommand{\tev}{\ensuremath{\,\mathrm{TeV}}}

\usepackage{amsmath}
\usepackage[colorlinks,citecolor=blue]{hyperref}
\hypersetup{colorlinks,citecolor=blue,linkcolor=blue,urlcolor=blue,}
\usepackage{graphicx} % Required for inserting images
\usepackage{subcaption}
\usepackage{multirow} % 引入 multirow 包以使用合并单元格

\usepackage{slashed}
\begin{document}
%\maketitle
\title{SKA Sensitivity to Potential Radio Emission from Dark Matter Annihilation in Ursa Major III}

\author{Peng-Long Zhang}
%\email{plzhang@ihep.ac.cn}
\affiliation{Institute of Particle and Nuclear Physics, Henan Normal University, Xinxiang 453007, China}
\affiliation{Key Laboratory of Particle Astrophysics, Institute of High Energy Physics, Chinese Academy of Sciences, Beijing 100049, China}

\author{Xiao-Jun Bi}
%\email{bixj@ihep.ac.cn}
\affiliation{Key Laboratory of Particle Astrophysics, Institute of High Energy Physics, Chinese Academy of Sciences, Beijing 100049, China}
\affiliation{School of Physical Sciences, University of Chinese Academy of Sciences, Beijing 100049, China}

\author{Qin Chang}
%\email{changqin@htu.edu.cn}
\affiliation{Institute of Particle and Nuclear Physics,
Henan Normal University, Xinxiang 453007, China}
\affiliation{Center for High Energy Physics, Henan Academy of Sciences, Zhengzhou 455004, China}

\author{Peng-Fei Yin}
%\email{yinpf@ihep.ac.cn}
\affiliation{Key Laboratory of Particle Astrophysics, Institute of High Energy Physics, Chinese Academy of Sciences, Beijing 100049, China}

\author{Yi Zhao}
%\email{aaaaaaaaa}
\affiliation{College of Physics and Materials Science, Tianjin Normal University, Tianjin 300387, China}

\begin{abstract}
The recently discovered stellar system, Ursa Major III/UNIONS 1, may be the faintest and densest dwarf spheroidal satellite galaxy of the Milky Way. 
Owing to its close proximity and substantial dark matter (DM) component, Ursa Major III  emerges as a highly promising target for DM indirect detection. It is known that electrons and positrons originating from DM annihilation can generate a broad radio spectrum through the processes of synchrotron radiation and inverse Compton scattering within galaxies. 
In this study, we investigate the potential of the Square Kilometre Array (SKA) in detecting radio signatures arising from DM annihilation in Ursa Major III over a 100 hour observation period. Our analysis indicates that the SKA has strong capabilities in detecting these signatures.
For instance, the SKA sensitivity to the DM annihilation cross section is estimated to reach $\mathcal{O}(10^{-30})-\mathcal{O}(10^{-28})\; \rm cm^{3} s^{-1}$ in the DM  mass range  from several GeV to $\sim100$ GeV for the $e^+e^-$ and $\mu^+\mu^-$ annihilation channels. The precise results are significantly influenced by various astrophysical factors, such as the strength of magnetic field, the diffusion coefficient, and the DM density profile in the dwarf galaxy. We discuss the impact of the uncertainties associated with these factors, and find that the SKA sensitivities have the potential to surpass the current constraints, even when considering these uncertainties.

\end{abstract}

%\date{\today}

%\begin{document}
\maketitle
%\flushbottom

\section{introduction}  

Astrophysical and cosmological observations have provided compelling evidence for the presence of non-baryonic cold dark matter (DM), comprising approximately 25\% of the total content of the Universe~\cite{Planck:2018vyg}. Among the various candidates proposed to elucidate the nature of DM, weakly-interacting massive particles (WIMPs) are particularly compelling. This is attributed to their strong theoretical motivation and their inherent ability to naturally explain the observed DM relic density~\cite{Bertone:2004pz}. The annihilation of WIMPs within galaxies and clusters has the potential to generate a variety of photon signatures.
Exploring these emissions provides a promising pathway towards uncovering the fundamental properties of DM.

The annihilation of WIMPs leads to the production of high energy primary particles that subsequently decay into detectable particles, including photons, electrons, and positrons.
These electrons and positrons traverse the interstellar medium and interact with their surroundings, giving rise to photon emissions through the mechanisms of synchrotron radiation and inverse Compton scattering (ICS). In the case of DM with masses above the GeV range, these synchrotron emissions can be detected by current or forthcoming radio telescopes ~\cite{Tyler:2002ux,Colafrancesco:2005ji,Colafrancesco:2006he,Natarajan:2013dsa,Spekkens:2013ik, Storm:2012ty, Regis:2014tga, Natarajan:2015hma, Colafrancesco:2014coa, Storm:2016bfw,McDaniel:2017ppt, Regis:2017oet, Cembranos:2019noa, Kar:2019cqo, Vollmann:2019boa, Kar:2020coz, Vollmann:2020gtu, Chen:2021rea,Guo:2022rqq,Vollmann:2020gtu, Gajovic:2023bsu, Regis:2023rpm, Wang:2023sxr}, such as GBT, VLA, LOFAR, ATCA, and FAST. 
On the other hand, for DM masses below $100\mev$,  the observation of emissions resulting from ICS within the radio spectrum remains feasible~\cite{Dutta:2020lqc}. 

The forthcoming Square Kilometre Array (SKA) radio telescope, renowned for its remarkable sensitivity and strong observational capabilities, is poised to explore the signatures produced by DM. 
With its broad frequency coverage spanning from 50 MHz to 50 GHz, the SKA stands as a powerful instrument for detecting emissions arising from DM annihilation, including synchrotron and ICS emissions. Many studies have explored the SKA capabilities in detecting potential DM signatures from a variety of astrophysical sources, such as Draco, Segue 1, A2199, Dragonfly 44, and Omega Centauri~\cite{Regis:2014tga, Cembranos:2019noa, Kar:2019cqo, Chen:2021rea, Wang:2023sxr, Dutta:2020lqc}. The results of these studies show that the  exceptional sensitivity of the SKA render it particularly effective at probing GeV scale DM through the detection of synchrotron emissions. 

Dwarf spheroidal galaxies (dSphs) are ideal targets for investigating photon signatures induced by WIMPs, owing to their advantageous characteristics, including high mass-to-light ratios, substantial DM densities, and minimal background interference~\cite{Baltz:1999ra, Bergstrom:2000pn, Evans:2003sc, Strigari:2007at, 2011PhRvL.107x1303G}. The recently discovered stellar system Ursa Major III/UNIONS 1 \cite{smith2023discovery, Errani:2023sgd}, hereafter referred to as Ursa Major III, emerges as such a particularly promising candidate. 
An analysis conducted in Ref.~\cite{Errani:2023sgd} elucidates that if Ursa Major III is a self-gravitating star cluster, it would face challenges in persisting over an extended period due to the gravitational  tidal forces exerted by the Milky Way. This circumstance strongly suggests the necessity of stabilizing Ursa Major III with a substantial amount of DM. Consequently, Ursa Major III may be the faintest and densest dSph associated with the Milky Way. 
The intrinsic stellar velocity dispersion of Ursa Major III is estimated to be $3.7^{+1.4}_{-1.0}~\rm km~s^{-1}$ \cite{smith2023discovery}, indicating a significant presence of DM within this galaxy. Given its substantial mass and proximity to Earth at approximately 10 kpc away, Ursa Major III may boast a considerable J-factor of $\sim 10^{21} \rm GeV^2 \rm cm^{-5}$~\cite{Errani:2023sgd,Crnogorcevic:2023ijs,Zhao:2024say}. Additionally, its high-galactic latitude positioning minimizes potential background interference originating from the Milky Way. 
These features enhance the appeal of Ursa Major III as a highly promising target for DM indirect detection \footnote{Nevertheless, it is important to note that the precise quantity of DM within Ursa Major III remains uncertain due to the limited kinematic data available. Excluding the member star with the largest velocity outlier, kinematic analyses suggest a significantly lower J-factor III~\cite{Errani:2023sgd,Zhao:2024say}, underscoring the necessity for further investigations to ascertain the amount of DM in Ursa Major.}.

In this study, we investigate the capabilities of the SKA in detecting the radio signatures generated by DM annihilation within Ursa Major III. By analyzing both synchrotron and ICS emissions, we aim to gain a comprehensive understanding of the the radio signatures produced by DM within this dSph. Our analysis considers the influence of varying magnetic field strengths and diffusion coefficients on these emissions. Furthermore, we address the uncertainties associated with the DM density profile of Ursa Major III, ensuring a thorough evaluations across a wide range of DM masses. We calculate the SKA sensitivities to the DM annihilation cross section
for three DM annihilation channels, including 
$e^+ e^-$, $\mu^+ \mu^-$, and $b \bar{b}$.

The paper is organized as follows. In Sec.~\ref{sec:SKA}, the SKA sensitivity 
is briefly described. We discuss the two phases of SKA1 and SKA2. 
In Sec.~\ref{sec:radio}, we recap the derivation of radio SED and show its 
variation when changing the magnetic field strength and diffusion coefficient.  
In Sec.~\ref{sec:result}, we present the sensitivity of DM annihilation cross section based on the minimal SKA1 flux density.    
We summarize our results in Sec.~\ref{sec:conclusion}.

\section{Dark matter density profile of Ursa Major III}
\label{sec:SKA}

In our investigation, we adopt the Navarro-Frenk-White density profile to characterize the DM distribution within Ursa Major III, which is expressed as
\begin{equation}
    \rho = \frac{\rho_{s}}{\frac{r}{r_{s}}  (1 + \frac{r}{r_{s}})^2},
    \label{eq:rho_NFW}
\end{equation}
where $\rho_s$ and $r_s$ represent the characteristic density and radius, respectively.
The DM density profile utilized in our study is derived from the analysis conducted in Ref.~\cite{Zhao:2024say}. In this section, we provide a a concise summary of the analysis methodology.

The dynamics of dSphs can be described by the Jeans equation, which is derived from the collisionless Boltzmann equation. Under the assumptions of spherical symmetry and a steady-state system with negligible rotational support, the second-order Jeans equation is given by~\cite{Courteau:2013cjm} 
\begin{eqnarray}\label{eq:jeans1}
\frac{1}{\nu(r)}\frac{d}{dr}[\nu(r)\sigma_r^2]+2\frac{\beta_{ani}(r)\sigma_r^2}{r}=-\frac{GM(r)}{r^2},
\end{eqnarray}
where $\nu(r)$ represents the three-dimensional stellar number density, $\sigma_r^2$ denotes the radial velocity dispersion of stars in the dSph, $\beta_{ani}(r)=1-\sigma_\theta^2/\sigma_r^2$ characterizes the stellar velocity anisotropy based on the ratio of tangential to radial velocity dispersions, $G$ is the gravitational constant, and $M(r)$ is the enclosed mass approximately determined by DM as $M(r)=4\pi\int_0^r\rho_{DM}(s)s^2ds$.
Considering that astrophysical observations typically provide only the two-dimensional projected stellar number density and the line-of-sight velocity dispersion, the solution to the Jeans equation, incorporating these observations, can be expressed as 
\begin{eqnarray}\label{eq:jeans2}
\sigma_p^2(R)=\frac{2}{I(R)}\int_R^{\infty}[1-\beta_{ani}(r)\frac{R^2}{r^2}]\frac{\nu(r)\sigma_r^2r}{\sqrt{r^2-R^2}}\mathrm{d}r,
\end{eqnarray}
where $\sigma_p(R)$ is the projected velocity dispersion corresponding to the projected radius $R$, and $I(R)$ denotes the projected light profile. The projected light profile of Ursa Major III has been estimated by assuming that its member stars follow an elliptical and exponential radial surface density profile~\cite{smith2023discovery}. Specifically, it can be expressed as $I(R)=I_{0}\exp(-R/r_{c})$ with $I_0=2.58\times 10^6\,\rm stars/\rm kpc^{2}$ and $r_c=1.52\times 10^{-3}\,\rm kpc$~\cite{Zhao:2024say}.

In Ref.~\cite{Zhao:2024say}, the Markov Chain Monte Carlo (MCMC) method, implemented using the GreAT toolkit within the CLUMPY package~\cite{Bonnivard:2015pia}, is employed to perform a Jeans analysis of Ursa Major III.
In this analysis, the parameters $\rho_s$ and $r_s$ characterizing the DM density profile, along with the anisotropy parameter $\beta_{ani}$ in Eq.~\ref{eq:jeans2}, are considered as free parameters in the MCMC analysis. The unbinned likelihood function is expressed as 
\begin{eqnarray}\label{eq:likelihood}
\mathcal{L}_{\mathrm{unbin}}=\prod_{i=1}^{N_{\mathrm{stars}}}\left(\frac{\exp\left(-\dfrac{1}{2}\dfrac{(v_i-\bar{v})^2}{\sigma_p^2(R_i)+\Delta_{v_i}^2}\right)}{\sqrt{2\pi\left[\sigma_p^2(R_i)+\Delta_{v_i}^2\right]}}\right)^{P_i},
\end{eqnarray}
where $v_i$ is the velocity of the $i$-th star, $\Delta_{v_i}$ is the uncertainty in the velocity measurement, $P_i$ indicates the membership probability, and $\bar{v}$ is the mean velocity of stars. The observational data for $R_i$, $v_i$, and $\Delta_{v_i}$ of the 11 member stars have been provided in Tab. 3 of Ref.~\cite{smith2023discovery}.
Through the MCMC analysis employing this methodology, more than 6000 DM density profiles are derived based on the posterior distribution, utilizing the available stellar kinematic data. The J factor within an angular radius of $0.5^\circ$ is calculated as $\log_{10}(J/\rm GeV^2 cm^{-5})= 21.4\pm 0.7$.

\section{Radio signatures from dark matter annihilation} 
\label{sec:radio}

In the context of DM annihilation, high energy $e^{\pm}$ can be produced directly or through cascade decays of other final states. As these particles travel through the galaxy, they have the potential to induce radio signals through the mechanisms of synchrotron radiation and ICS. In this study, the RX-DMFIT package~\cite{McDaniel:2017ppt} is utilized to compute the radio flux density originating from DM annihilation. In this section, we outline the computational methodology utilized for radio signatures.

In order to derive the radio signatures, it is crucial to obtain the spectrum of $e^{\pm}$ particles subsequent to undergoing propagation processes. 
The equilibrium $e^{\pm}$ spectrum can be obtained by solving the diffusion equation 
\begin{equation}
    \frac{\partial}{\partial t} \frac{\partial n_e}{\partial E}   =
    \nabla\left[ D(E,\textbf{r}) \nabla \frac{\partial n_e}{\partial E}\right] 
    +\frac{\partial }{\partial E}\left[b(E,\textbf{r})\frac{\partial n_e}{\partial E}\right] + Q(E,\textbf{r}),
    \label{eq:diffep}
\end{equation}
where $\frac{\partial n_e}{\partial E}$ represents the equilibrium electron density, $Q(E,\textbf{r})$ denotes the source term, $D(E,\textbf{r})$
is the diffusion coefficient, and $b(E,\textbf{r})$ is the energy loss term experienced by the $e^\pm$ particles during propagation. 
The source term, encompassing the contributions from DM annihilation, is given by
\begin{equation}
Q(E,\textbf{r}) =
 \frac{\rho(r)^{2}}{2 m^{2}_{\chi}} \sv \frac{dN_e}{dE},
\label{source_function}
\end{equation}
where $\rho$ represents the DM density, $\sv$ is the velocity averaged DM annihilation cross section, $m_\chi$ is the DM mass, and $dN_e/dE$ is the initial $e^{\pm}$ injection spectrum per WIMP annihilation event,
obtained using the DarkSUSY package.

High energy $e^{\pm}$ particles undergo complicated propagation effects within the galaxy. Here, we briefly provide the description of the diffusion coefficient and the energy loss term, that are crucial in modeling the propagation of these particles.
The diffusion coefficient can be given by the expression: 
\begin{equation}
D(E)=D_0 (E/E_0)^\gamma, 
\end{equation}
where $D_0$ represents the diffusion coefficient with the reference energy $E_0$ in the GeV scale.
In this study, we adopt $\gamma=0.3$ corresponding to the Kolmogorov turbulence scenario. The determination of the diffusion coefficient relies on measurements of cosmic-ray secondaries, such as the B/C ratio. In the Milky Way, the value of $D_0$ is determined to be $\mathcal{O}(10^{28}) { \rm cm^2 s^{-1}}$ at the GeV scale.  
The energy loss term $b(E, \mathbf{r})$ includes the effects of various processes, including synchrotron radiation, ICS, Coulomb interaction, and bremsstrahlung. It can be expressed as 
\begin{equation}
\begin{aligned}
b(E, \mathbf{r}) &=b_{\rm IC}(E)+b_{\rm syn.}(E, \mathbf{r})+b_{\text {coul. }}(E)+b_{\text {brem. }}(E) \\
&=b_{\rm IC}^{0} E^{2}+b_{\rm syn.}^{0} B^{2}(r) E^{2} +
b_{\text {coul. }}^{0} n_{e}\left[1+\log \left(\frac{E / m_{e}}{n_{e}}\right) / 75\right]\\
& +b_{\text {brem. }}^{0} n_{e}\left[\log \left(\frac{E / m_{e}}{n_{e}}\right)+0.36\right], 
\end{aligned}
\end{equation}
where $n_e$ represents the average thermal electron density in units of $\rm cm^{-3}$, 
assumed to be  
$10^{-6}$ in this study, and $B$ represents the magnetic field strength in units of $\mu \rm G$. 
The energy loss coefficients $b_{\rm IC}^{0}$, $ b_{\rm syn.}^{0}$, $b_{\rm coul. }^{0}$, and $b_{\rm brem. }^{0}$ are set as 0.25, 0.0254, 6.13, and $1.51\times 10^{-16}\rm GeV~s^{-1}$, respectively~\cite{Colafrancesco:2006he, Colafrancesco:2014coa}. 

The solution in a state of equilibrium to Eq.\eqref{eq:diffep} can be obtained by setting the temporal evolution of the $e^\pm$ spectrum to zero.
The distribution of electrons is given by
\begin{equation}
\label{eq:dndeEq}
\frac{\partial n_e}{\partial E } = 
\frac{1}{b(E,\textbf{r})}\int_E^{m_{\chi}} dE' G\left(\textbf{r}, E, E'\right)Q(E, \textbf{r}),
\end{equation}
where $G\left(\textbf{r}, E, E'\right)$ denotes the Green function including the diffusion and energy loss effects. A detailed elucidation of this solution can be found in Ref.~\cite{Colafrancesco:2005ji}. The Green function $G\left(\textbf{r}, E, E'\right)$ determined with the free escape boundary condition at the radius of the diffusion region $r_h$ is expressed as 
\begin{multline}
G(\textbf{r},  E,  E') = \frac{1}{\sqrt{4\pi \lambda^2 }}\sum_{n = -\infty}^{\infty} \left(-1\right)^n\left\{\int_0^{r_h}dr' \frac{r'}{r_n} \left(\frac{ \rho_{\chi}(r') }{\rho_{\chi}(r)}\right)^2 \right.\\ \left.
 \times\left[
\exp\left(-\frac{ (r' - r_n)^2 }{4 \lambda^2}	\right)	- 	\exp\left(-\frac{ (r' + r_n)^2 }{4 \lambda^2}	\right)	\right] \right\}, 
\end{multline} 
where $r_n=(-1)^nr+2nr_h$.  The parameter $\lambda$ represents the mean energy-loss 
distance of $e^\pm$ with the initial energy $E'$ and the interaction energy $E$, which is given by
\begin{equation}
\label{eq:ve}
\lambda^2 (E, E') = \int_E^{E'} d\bar{E}\frac{D(\bar{E})}{b(\bar{E})}.\end{equation}

Utilizing the solution provided by Eq.~\ref{eq:dndeEq}, 
the integrated radio flux density over the solid angle, surrounding the line of sight from the Earth to the dSph's center, can be expressed as
\begin{eqnarray}\label{eq:SED}
S(\nu) &=& \int_{\Omega} d\Omega \int_{l.o.s.} \frac{dl}{4\pi} (j_{syn}(\nu, l) + j_{IC}(\nu, l)) \nonumber\\ 
&=& \int_{\Omega} d\Omega \int_{l.o.s.} \frac{dl}{4\pi} \int_{m_e}^{m_\chi} dE \times 2\frac{dn_e}{dE}(E, r) ( P_{syn}(\nu, E, r) + P_{IC}(\nu, E)) ,
\end{eqnarray}
where $r$ denotes the distance between the emitting point and the center of the dSph, $l$ represents the distance between the emitting point and observer, the factor of 2 is included to account for the equal contributions of electrons and positrons, and $P_{syn}$ and $P_{IC}$ correspond to the synchrotron and ICS powers, respectively.
The detailed formulae of  $P_{syn}$ and $P_{IC}$
can be found in the literature, e.g.~\cite{Colafrancesco:2005ji, McDaniel:2017ppt}.

DM annihilation has the potential to generate significant radio signatures.  While no definitive radio signatures induced by DM have been observed to date, the imposition of stringent constraints can be achieved through the utilization of radio results. However, the reliability of such calculations is affected by significant uncertainties arising from various astrophysical factors, thereby diminishing the robustness of the derived constraints. When considering radio signals originating from DM annihilation within dSphs, the primary sources of uncertainty pertain to the poorly understood magnetic fields and diffusion processes within these astrophysical environments. The in-depth discussions on these issues can be found in Ref.~\cite{Regis:2014tga, Regis:2014koa}. 

The strength of the magnetic field directly plays a pivotal role in determining the intensity of synchrotron radiation. Dynamic processes in galaxies like supernova explosions are expected to be potent sources for generating magnetic fields, thereby establishing a correlation between the magnetic field strength and the star formation rate within galaxies~\cite{Regis:2014koa}. These magnetic fields, originating from dynamic astrophysical phenomena, have the capacity to persist over extended time scales despite undergoing dissipation processes. Observational data have substantiated the existence of magnetic fields in certain dwarf galaxies~\cite{Chyzy:2011sw}. For instance, a magnetic field of $\sim 4.3~\mu\rm G$ has been detected in the Large Magellanic Cloud~\cite{Gaensler:2005qj}. Therefore, it is plausible to infer that dSphs harbor magnetic fields of the order of $\mathcal{O}(0.1)-\mathcal{O}(1)\mu\rm G$. These magnetic fields are hypothesized to exhibit a radial dependence characterized by the functional form $B(r) = B_{c} \exp(-r/r_{c})$,
where $B_c$ denotes the central magnetic field strength and $r_c$ is taken to be
the half-light radius of the dSph.
While the magnetic field strength within dSphs may be diminished in scenarios of limited star formation activity, charged particles traversing these galaxies can still be influenced by magnetic fields originating from the Milky Way. Galactic outflows are capable of magnetizing the vicinity surrounding dSphs \cite{Regis:2014koa}. Given the proximity of Ursa Major III to the Galactic center, it is plausible that this magnetic field exerts an impact on charged particles within the dSph. In the context of the limited diffusion region under consideration, the strength of this magnetic field can be approximated as a constant denoted by $B=B_0$.

The flux of radio signatures originating from dSphs is dependent on the $e^\pm$ density particles within these astrophysical environments.  The equilibrium $e^\pm$ spectrum is shaped by characteristic time scales governed qualitatively by  $\frac{\partial n_e}{\partial E} \propto Q\cdot min\{\tau_{loss}, \tau_{diff}\}$, where $\tau_{loss} \sim E/b(E)$ and $\tau_{diff} \sim R^2/D(E)$ represent the energy loss and diffusion time scales, respectively. In the context of dSphs, where the size is relatively small, it is expected that $\tau_{diff}<\tau_{loss}$ for $e^\pm$ with low energies. This implies that the flux of the low energy $e^\pm$ in the dSph is predominantly influenced by the diffusion effect.
While the diffusion coefficient within the Milky Way is determined to be $D_0 \sim \mathcal{O}(10^{28}) \rm cm^2 \rm s^{-1}$, the diffusion coefficient in dSphs remains ambiguous. The rapid diffusion of charged particles within certain dSphs is anticipated due to the significant suppression of astrophysical processes and the absence of accompanying turbulence. In an extreme scenario characterized by free streaming, charged particles directly escape from the dSph due to an exceedingly large diffusion coefficient. However, it is important to note that the charged particles generated by DM annihilation also have the potential to induce magnetic irregularities, thereby leading to non-negligible diffusion effects within dSphs~\cite{Regis:2014tga, Regis:2014koa}. A detailed exploration of this mechanism within dSphs can be found in Ref.~\cite{Regis:2023rpm}.

\section{The radio spectrum of dark matter annihilation in Ursa Major III}
\label{sec:SED}

In this section, we analyze the spectral energy distribution (SED) of radio emissions generated by DM annihilation within Ursa Major III. We consider a DM annihilation cross section of  $\sv=10^{-28}$ $\mathrm{cm^3 s^{-1}}$ for the $e^+e^-$ channel, and adopt a density profile characterized by $\rho_{s} = 7.59~\mathrm{GeVcm^{-3}}$ and $r_{s} = 1~\mathrm{kpc}$, resulting in a typical J-factor of $10^{21.3}~\rm GeV^2 cm^{-5}$. Furthermore, we investigate the impact of astrophysical factors, including the diffusion parameters and the strength of the magnetic field, on the SED of radio emissions originating from DM annihilation in Ursa Major III.

In the specified benchmark scenarios, we examine two distinct astrophysical parameter sets. We consider a constant magnetic field within the dSph, accounting for the potential influence of the Milky Way.
The first parameter set comprises a moderate constant magnetic field strength of $B_0=1 \mu\rm G$ and an ordinary diffusion coefficient of $D_0=3\times 10^{28} \rm cm^2 s^{-1}$, reflecting an optimistic scenario. The second parameter set involves a lower magnetic field strength of $B_0=0.1 \mu\rm G$ and a higher diffusion coefficient of $D_0= 10^{30} \rm cm^2 s^{-1}$, representing a conservative scenario.
The radius of the diffusion region, denoted as $r_h$, is assumed to be 10 times the half-light radius of Ursa Major III, with $r_c = 3 \rm pc$.

\begin{figure}[htbp]
\centering
\begin{subfigure}{0.49\textwidth}
    \centering
    \includegraphics[width=\textwidth]{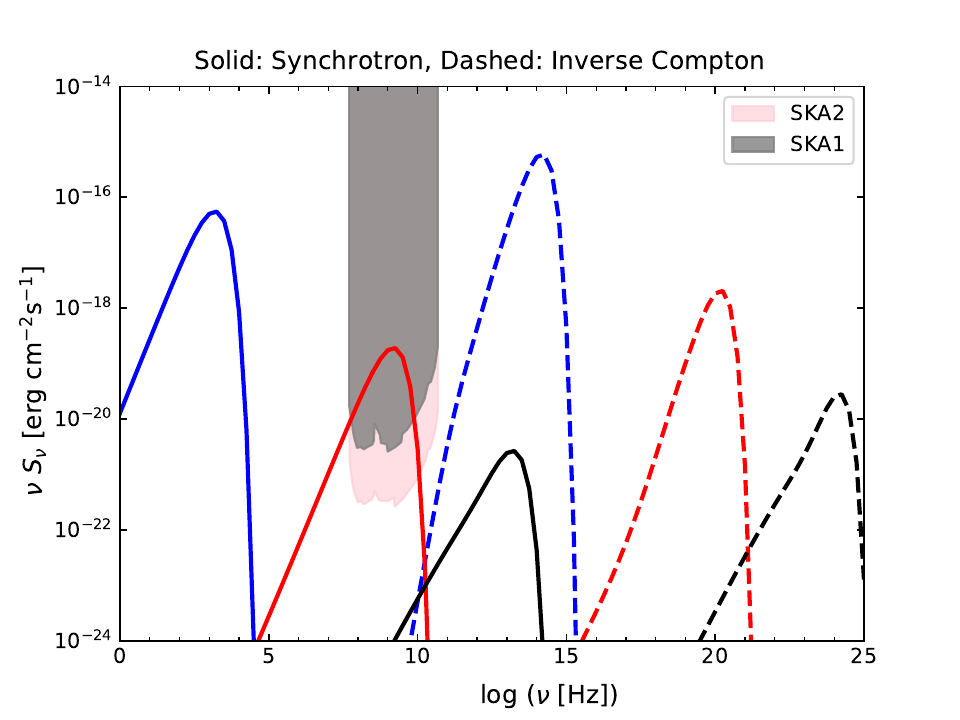}
    \subcaption{(a)D$_{0}$= $3 \times 10^{28}\rm cm^2\rm s^{-1}$, B$_{0}$= $1~\rm \mu G$}
    \label{fig:SED_mx_subfig_a}
\end{subfigure}
\hfill
\begin{subfigure}{0.49\textwidth}
    \centering
    \includegraphics[width=\textwidth]{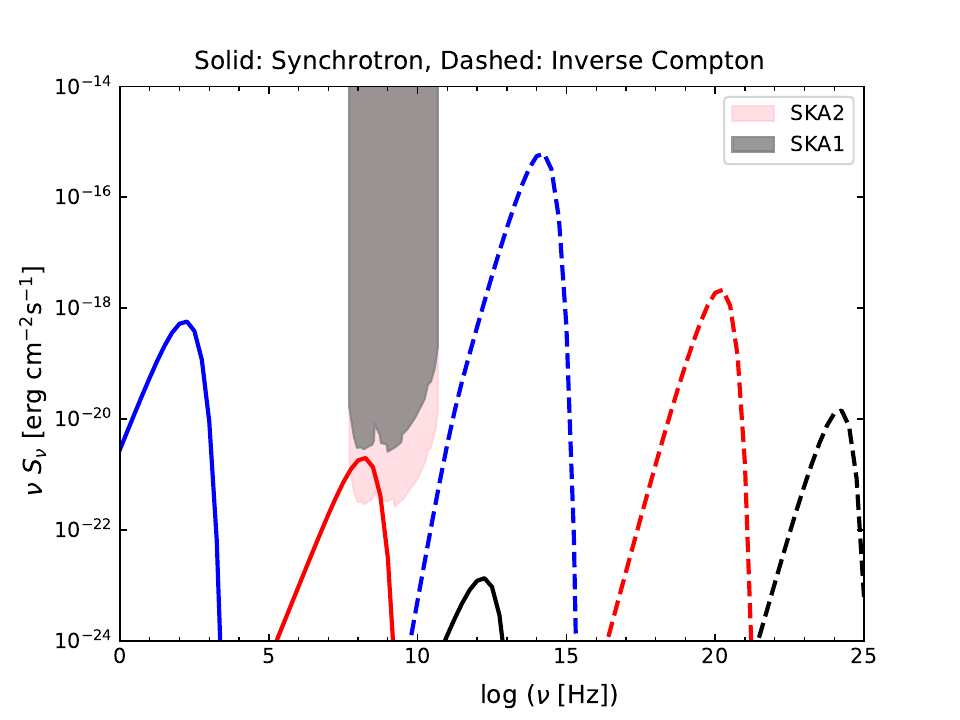}
    \subcaption{(b)D$_{0}$= $10^{30}\rm cm^2\rm s^{-1}$, B$_{0}$= $0.1~\rm \mu G$}
    \label{fig:SED_mx_subfig_b}
\end{subfigure}
\caption{ 
The SEDs of synchrotron radiation (solid lines) and ICS (dashed lines) emissions for various choices of DM mass: 10 MeV (blue), 10 GeV (red), and 1 TeV (black). The left and right panels corresponding to the optimistic and conservative scenarios, respectively.
}
\label{Fig:SED_mx}
\end{figure}

In Fig.~\ref{Fig:SED_mx}, we display the SED for three distinct DM masses: 10$\mev$, 10$\gev$, and 1 $\tev$ in the two scenarios. 
The projected sensitivities of SKA1 and SKA2 are depicted by the grey and pink contours, respectively. 
The contributions of synchrotron radiation and ICS are represented by the solid and dashed lines, respectively. 
The results in Fig.~\ref{Fig:SED_mx} show that the ICS contribution prevails at higher frequencies, while synchrotron radiation dominates at lower frequencies. For a large DM mass of 1 TeV, both the peak frequencies of synchrotron and ICS spectra exceed the detection range of the SKA, rendering them undetectable. Only the contributions of ICS for $m_\chi = 10~\rm MeV$ and synchrotron radiation for $m_\chi = 10~\rm GeV$ may fall within the detectable frequency range of the SKA.

\begin{figure}[htbp]
\centering
\begin{subfigure}{0.49\textwidth}
    \centering
    \includegraphics[width=\textwidth]{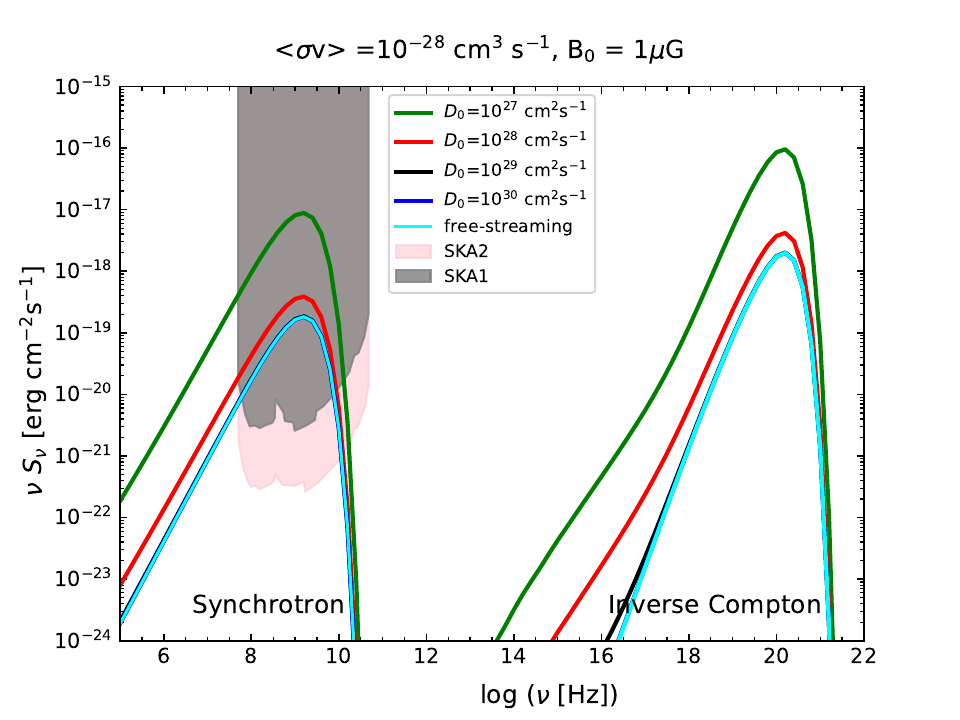}
    \subcaption{(a) B$_{0}$= $1~\rm \mu G$}
    \label{fig:dif_D0_subfig_a}
\end{subfigure}
\hfill
\begin{subfigure}{0.49\textwidth}
    \centering
    \includegraphics[width=\textwidth]{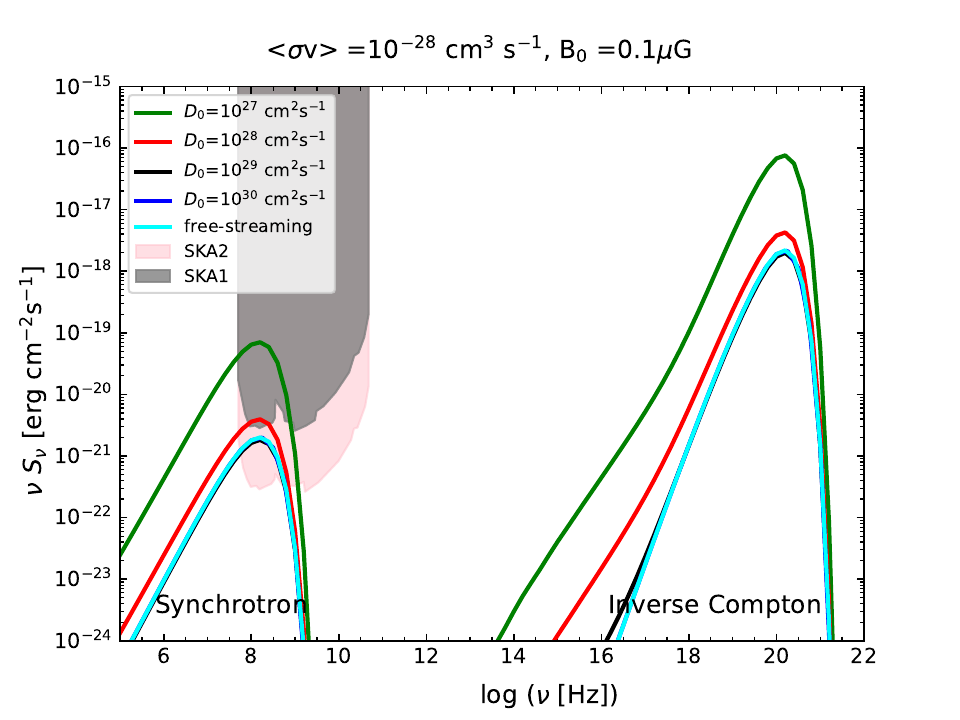}
    \subcaption{(b) B$_{0}$= $0.1~\rm \mu G$}
    \label{fig:dif_D0_subfig_b}
\end{subfigure}
\caption{ 
The SEDs for varying diffusion coefficients ($D_0= 10^{27}, 10^{28}, 10^{29}, 10^{30} \rm cm^2\rm s^{-1}$) alongside the free-streaming scenario. The DM mass is taken to be 10 GeV. The left and right panels correspond to the results for $B_0= 1 \mu\rm G$ and $0.1 \mu\rm G$, respectively.
}
\label{Fig:IC_syn_dif_D0}
\end{figure}

In our investigation of the influences of astrophysical factors on the SED, we focus on the complexities associated with the diffusion effect and magnetic fields within the dSph. 
The SEDs corresponding to different diffusion coefficients  ($D_0= 10^{27}, 10^{28}, 10^{29}, 10^{30} \rm cm^2\rm s^{-1}$) alongside the free-streaming scenario are presented in Fig.~\ref{Fig:IC_syn_dif_D0}. We consider  the diffusion of $e^\pm$ particles in a constant magnetic field of $0.1 \mu$G or $1 \mu$G. The SEDs exhibit remarkable similarity for diffusion coefficients $D_0 \gtrsim 10^{29} \rm m^2 \rm s^{-1}$, primarily due to the rapid escape of $e^\pm$ particles from the small diffusion region with $r_h=0.03$ kpc.

\begin{figure}[htbp]
\centering
\begin{subfigure}{0.49\textwidth}
    \centering
    \includegraphics[width=\textwidth]{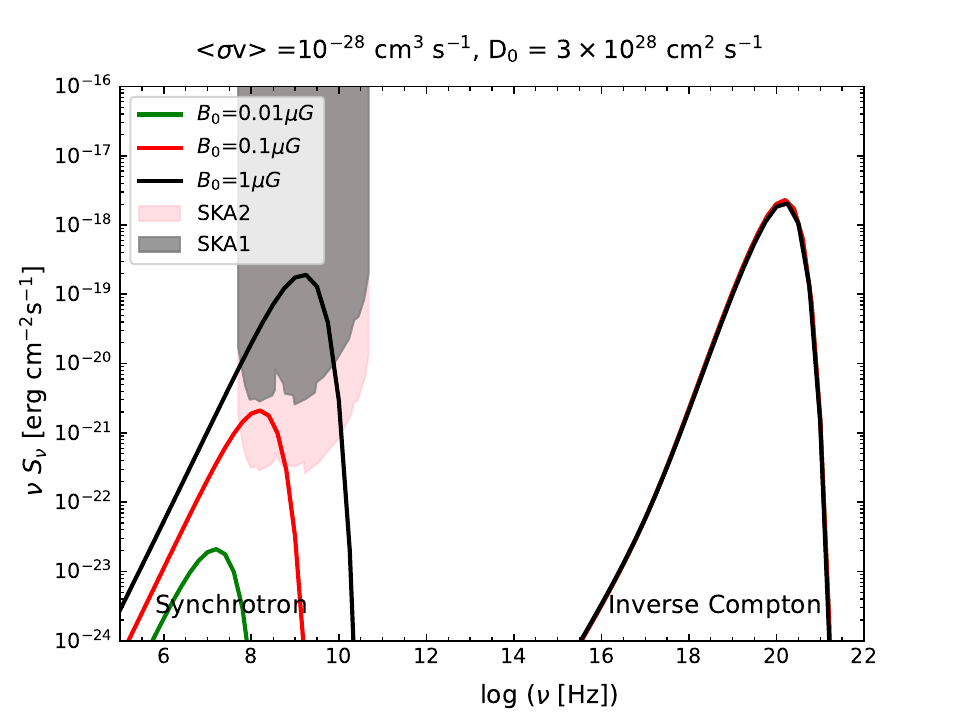}
    \subcaption{(a)D$_{0}$= $3 \times 10^{28}\rm cm^2\rm s^{-1}$}
    \label{fig:dif_B0const_subfig_a}
\end{subfigure}
\hfill
\begin{subfigure}{0.49\textwidth}
    \centering
    \includegraphics[width=\textwidth]{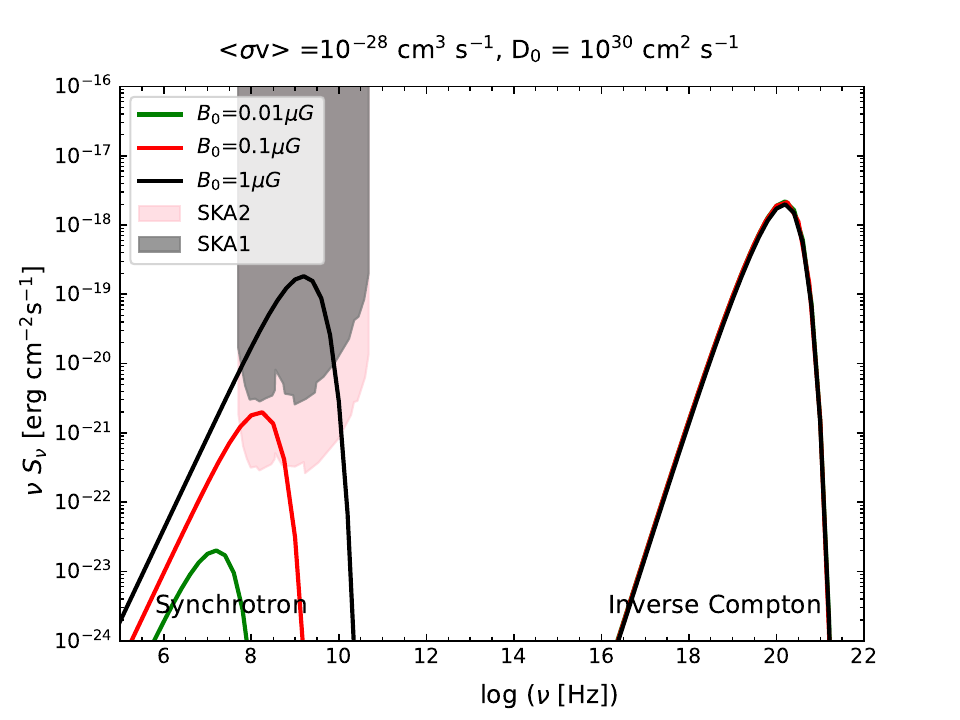}
    \subcaption{(b) D$_{0}$= $10^{30}\rm cm^2\rm s^{-1}$}
    \label{fig:dif_B0const_subfig_b}
\end{subfigure}
\caption{ 
The SEDs for varying magnetic field strengths ($B_0= 0.01, 0.1, 1 \mu\rm G$) alongside the free-streaming scenario. The DM mass is taken to be 10 GeV. The left and right panels correspond to the results for $D_0= 3\times 10^{28} \rm cm^2 s^{-1}$ and $10^{30} \rm cm^2 s^{-1}$, respectively.
}
\label{Fig:dif_B0const}
\end{figure}

In Fig.\ref{Fig:dif_B0const}, we investigate the impact of varying magnetic field strengths on the SED. Specifically, we examine SEDs corresponding to different values of  $B_0$ as $0.01$, $0.1$, and $1 \mu\rm G$. 
It is evident that the value of $B_0$ exerts a significant influence on the synchrotron radiation contributions. As the magnetic field strength increases, the flux density of the synchrotron SED significantly intensifies, accompanied by a shift in its peak towards higher frequencies. This spectral shift can be understood by the peak frequency $\nu_p \sim (4.7 \rm MHz) (B/\mu\rm G) (E/GeV)^2$ estimated within the monochromatic approximation in Ref.~\cite{Colafrancesco:2005ji}. In contrast, the SEDs stemming from the ICS process remain largely unaffected by variations in the magnetic field strength. The left and right panels of Fig.\ref{Fig:dif_B0const} represent the results for $D_0=3 \times 10^{28} \rm cm^2 s^{-1}$ and $10^{30} \rm cm^2 s^{-1}$, respectively, revealing that the diffusion coefficient exerts minor influence on the SEDs.

\begin{figure}[htbp]
\centering
\begin{subfigure}{0.49\textwidth}
    \centering
    \includegraphics[width=\textwidth]{
    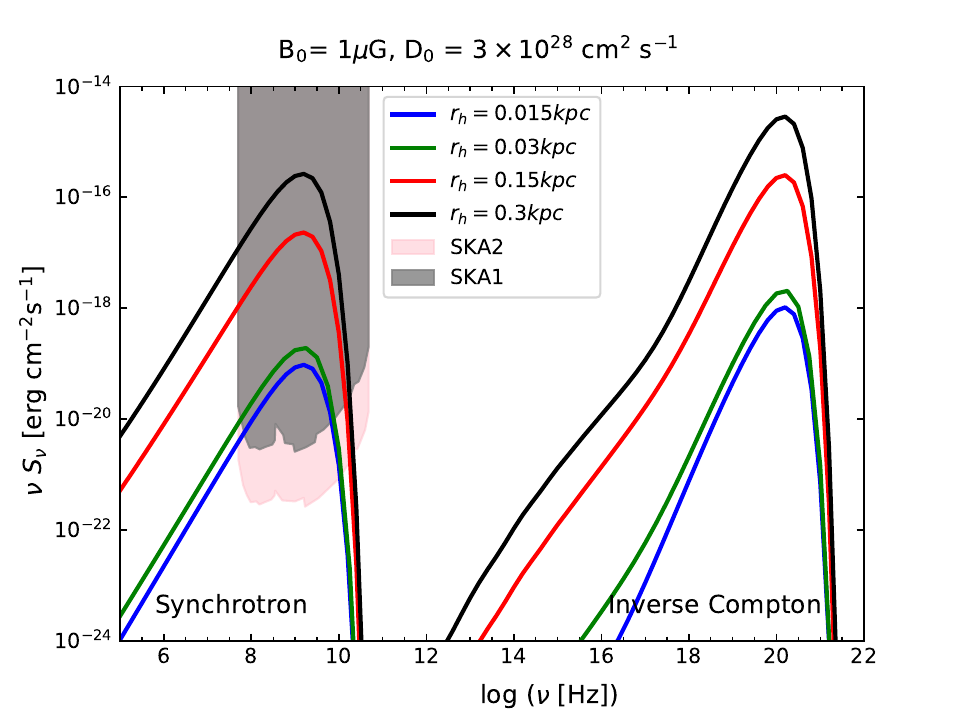}
    \subcaption{(a)
    D$_{0}$= $3 \times 10^{28}\rm cm^2\rm s^{-1}$, B$_{0}$= $1~\rm \mu G$
    }
    \label{fig:SED_rh_subfig_a}
\end{subfigure}
\hfill
\begin{subfigure}{0.49\textwidth}
    \centering
    \includegraphics[width=\textwidth]{
    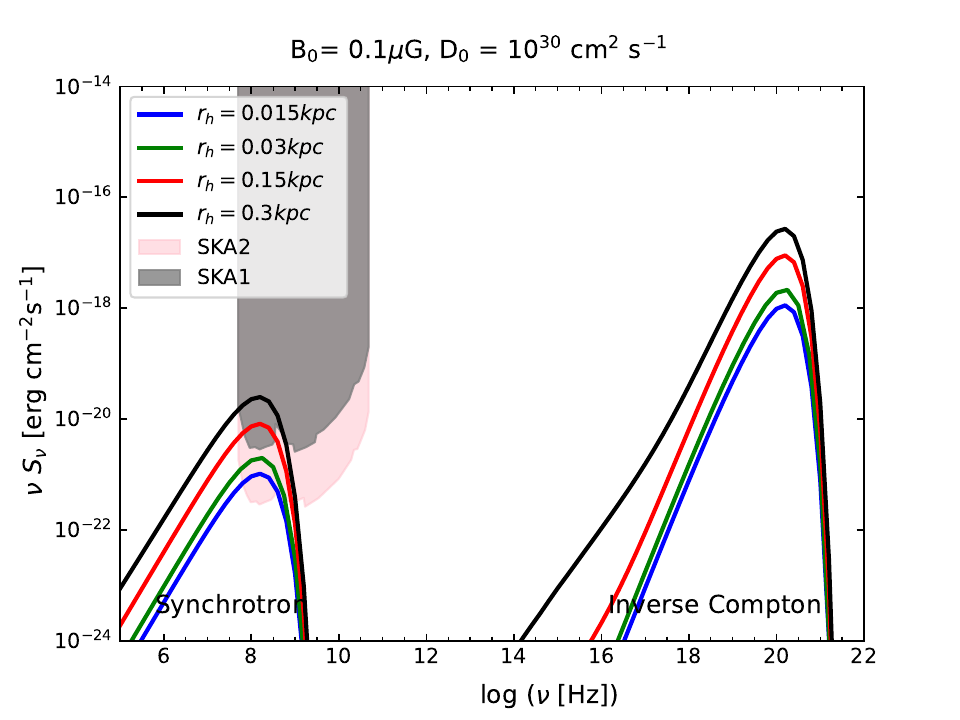}
    \subcaption{(b)
    D$_{0}$= $10^{30}\rm cm^2\rm s^{-1}$, B$_{0}$= $0.1~\rm \mu G$
    }
    \label{fig:SED_rh_subfig_b}
\end{subfigure}
\caption{ 
The SEDs for $m_\chi=10$ GeV and the different values of the radius of the diffusion region ($r_h=$ 0.015, 0.03, 0.15, and 0.3kpc). The left and right panels correspond to the results in the optimistic and conservative scenarios, respectively.
}
\label{Fig:SED_rh}
\end{figure}

In the aforementioned analysis, the radius of the diffusion region $r_h$ has been set at $0.03$ kpc.
The SEDs corresponding to various values of $r_h$ under the two scenarios are presented in Fig.~\ref{Fig:SED_rh}. Given that $r_h$ plays an important role in determining the diffusion time of $e^\pm$ in the dSph, it exerts a profound impact on the flux of $e^\pm$. Consequently, significant variations in the SEDs manifest with changes in $r_h$. As illustrated in Fig.~\ref{Fig:SED_rh}, a conspicuous trend emerges where an increase in $r_h$ correlates with a corresponding rise in the SED magnitude. 
Moreover, it is evident that for a high magnetic field strength and a low diffusion coefficient, an increase in $r_h$ results in a substantial enhancement of the synchrotron radiation flux. These behaviors underscore the intricate relationship between $r_h$ and the resulting SEDs.

%\begin{figure}[htbp]
%\centering
%\includegraphics[width=0.49\textwidth]{draw_SED_IC_syn_D0_B0const_rh_01.pdf}
%\hfill
%\caption{The SEDs for different values of $r_h$: 0.015, 0.03, 0.15, 0.3, and 0.6 kpc. The solid and dashed lines correspond to the ICs and synchrotron emissions for DM masses of $m_\chi = 10~\rm MeV$ and $m_\chi = 10~\rm GeV$, respectively. The remaining parameters are consistent with those ultilized in Fig.~\ref{Fig:SED_mx}.   
%}
%\label{Fig:SED_rh}
%\end{figure}

%%%%%%%%%%%%%%%%%%%%%%%%%%%%%%%%%%%%%%%%%%%%%%%%%%%%
\section{SKA sensitivities}
\label{sec:result}
%%%%%%%%%%%%%%%%%%%%%%%%%%%%%%%%%%%%%%%%%%%%%%%%%%%%

The sensitivity to DM annihilation signatures can be determined through the flux density measurements achievable with the telescope. The minimum detectable flux of the SKA telescope can be quantified by the expression~\cite{Braun:2019gdo}: 
\begin{eqnarray}
 S_{\text{min}}=\frac{2 k_{b}S_{D} T_{\text{sys}}}{\eta_{s}A_{e}(\eta_{\text{pol}} t \Delta\nu)^{1/2}}\, ,
\label{S_min}
\end{eqnarray}
where $k_{b}$ denotes the Boltzmann constant, $S_{D}$ is a degradation factor, $\eta_s$ represents the system efficiency, $\eta_{\rm pol}$ is the number of polarization states, $t$ is the total observation time, $A_{e}$ is the effective collecting area, $T_{sys}$ denotes the total system noise temperature, and $\Delta\nu$ is the channel bandwidth. 
In our analysis, we consider an integration time of 100 hours, covering the detection frequency range of SKA from 50 MHz to 15.4 GHz. We utilize the SKA sensitivity calculator \footnote{https://www.skao.int/en/science-users/ska-tools/493/ska-sensitivity-calculators} based on \cite{Sokolowski_2022} to compute the sensitivities for the SKA low band ($0.05-0.35$ GHz), and SKA mid bands B1 ($0.35-1.05$ GHz), B2 ($0.95-1.76$ GHz), B5a ($4.6-8.5$ GHz), and B5b ($8.3-15.4$ GHz). Subsequently, we project the SKA sensitivities to the DM annihilation cross section.

From the expressions of the synchrotron and ICS contributions given by Eq.~\eqref{eq:SED} and the DM source term represented by Eq.~\eqref{source_function}, the flux density can be reformulated as:
\begin{equation}
S = \frac{\langle \sigma v \rangle} {m^{2}_{\chi}} f(\nu),
\label{eq:flux1}
\end{equation}
where $f(\nu)$ encompass information independent of the DM mass and annihilation cross section. The sensitivity to the DM annihilation cross section can be given by:
\begin{equation}
\langle \sigma v \rangle = \frac{S_{min}}{f} m^{2}_{\chi},
\label{eq:sigma_v}
\end{equation}
where $S_{min}$ denotes the flux density sensitivity of the SKA.

In Fig.~\ref{Fig:sigma_v}, we illustrate the SKA sensitivity to the DM annihilation cross section, employing the SEDs of ICS and synchrotron emissions within the optimistic scenario for three annihilation channels: $e^+e^-$, $\mu^+\mu^-$, and $b\bar{b}$. 
In order to account for the uncertainties associated with the DM density distribution in Ursa Major III, we consider all DM density profiles obtained from the Jeans analysis and calculate the corresponding sensitivity for each density profile. The red lines in the plot denote the median sensitivities, while the yellow and green bands represent the 68\% and 95\% containment bands, respectively. 

\begin{figure}[htbp]
\centering
\begin{subfigure}{0.48\textwidth}
    \centering
    \includegraphics[width=\textwidth]{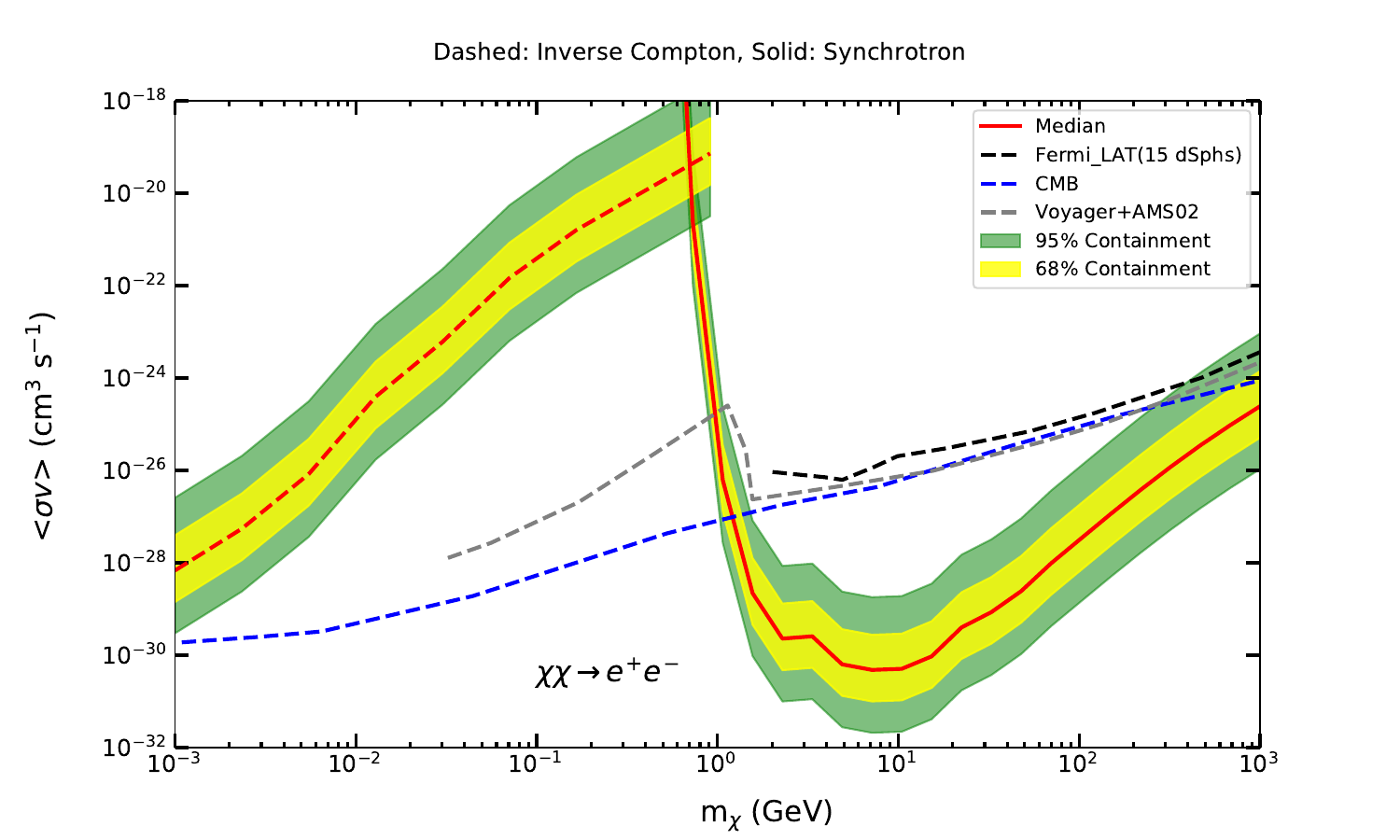}
    \subcaption{(a) $e^{+}e^{-}$ channel}
    \label{fig:ee_sigmav_p_subfig_a}
\end{subfigure}
\hfill
\begin{subfigure}{0.48\textwidth}
    \centering
    \includegraphics[width=\textwidth]{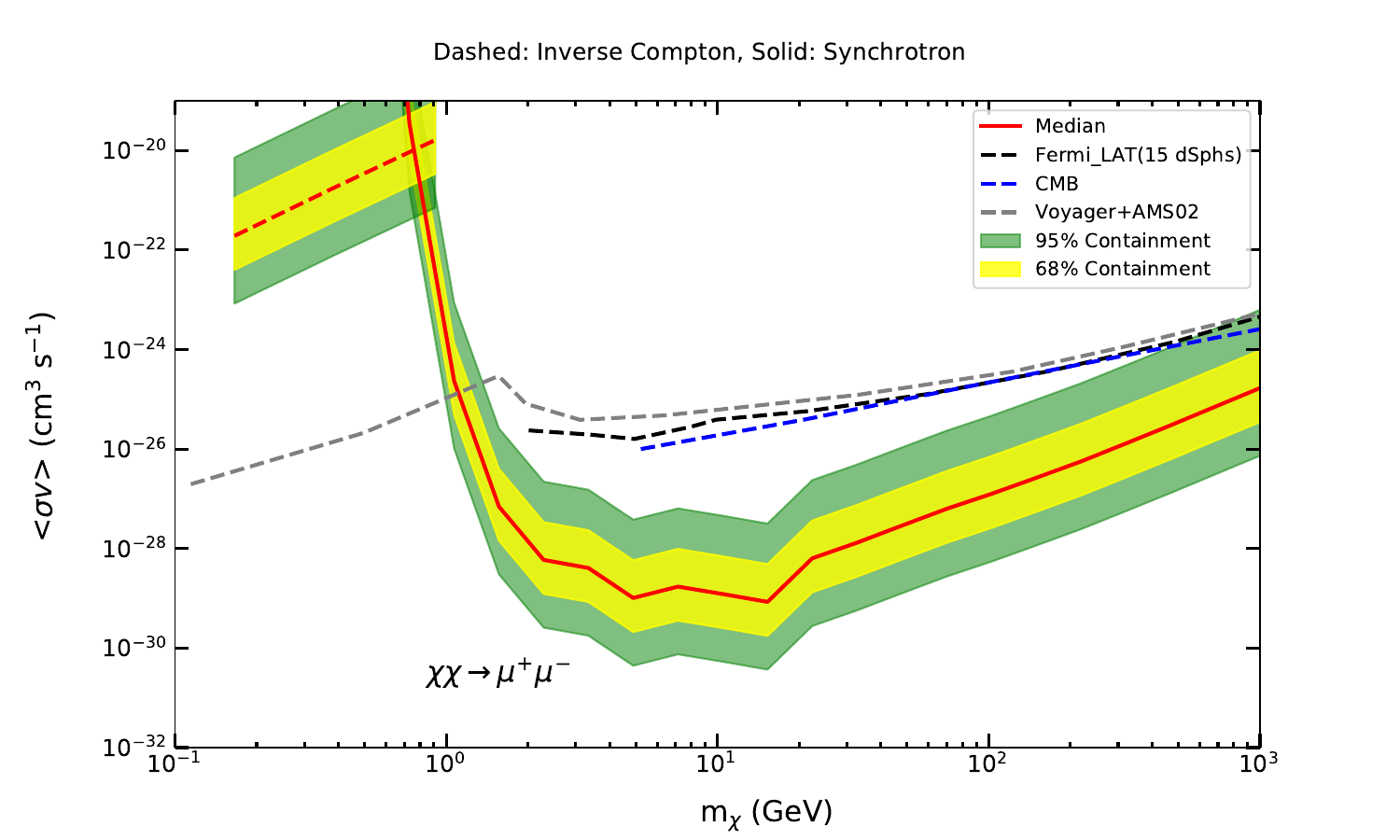}
    \subcaption{(b) $\mu^{+}\mu^{-}$ channel}
    \label{fig:mumu_sigamv_p_subfig_b}
\end{subfigure}
\hfill
\begin{subfigure}{0.48\textwidth}
    \centering
    \includegraphics[width=\textwidth]{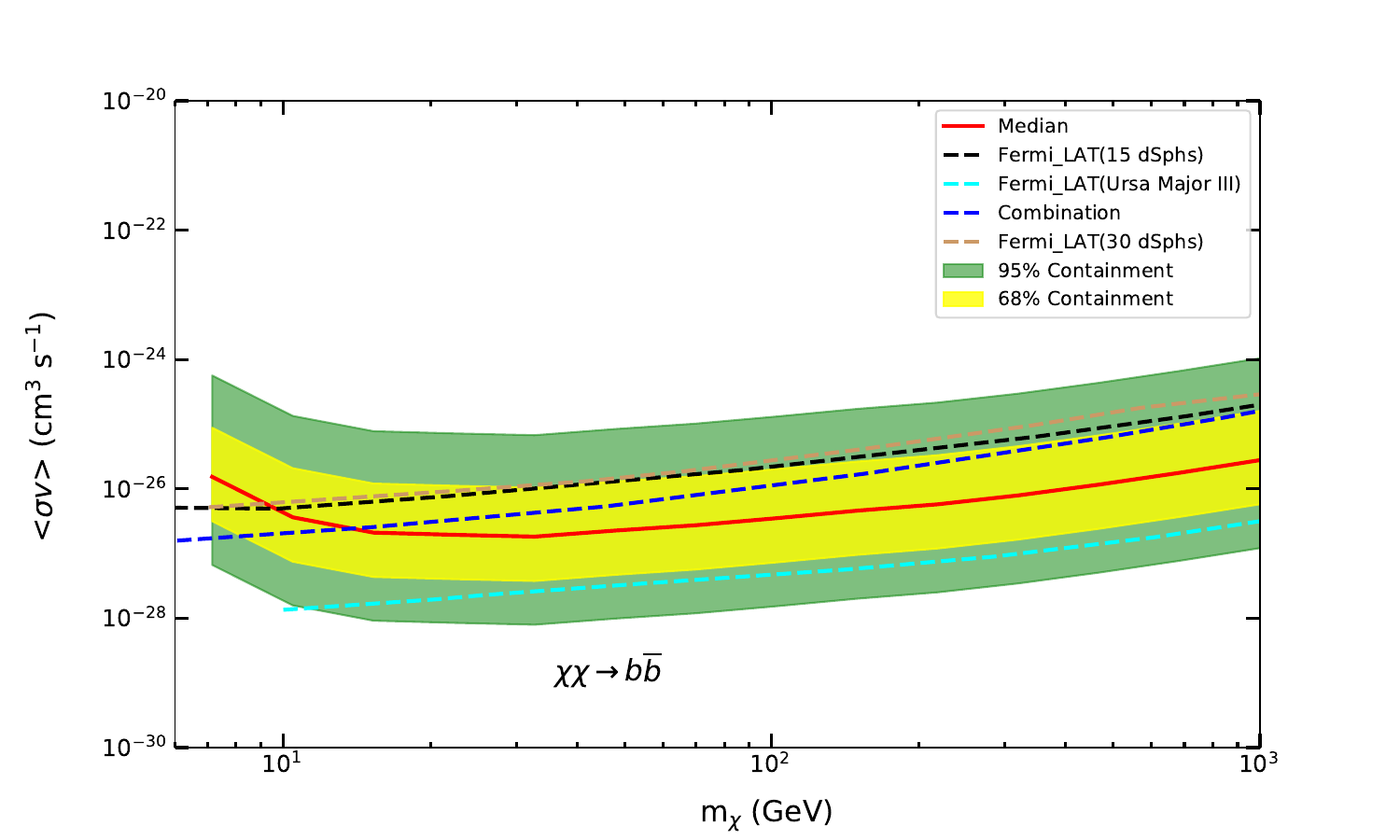}
    \subcaption{(c) $b\overline{b}$ channel }
    \label{fig:bb_bar_sigamv_p_subfig_c}
\end{subfigure}
\caption{The median sensitivities of the SKA1 (red solid lines), along with their 68\% (yellow bands) and 95\% (green bands) containment bands accounting for uncertainties of the DM density profile in Ursa Major III in the optimistic scenario for the 
(a) $e^{+}e^{-}$, (b) $\mu^{+}\mu^{-}$, and (c) $b\bar{b}$ annihilation channels. The astrophysical parameters considered are
$B_{0}= 1\mu$G, $D_{0} = 3\times 10^{28} \rm cm^2 s^{-1}$, and $r_{h}=0.03$ kpc. The constraints obtained from other observations are also shown for comparison.
}
\label{Fig:sigma_v_positive}
\end{figure}

The analysis depicted in Figure~\ref{Fig:sigma_v_positive} reveals that the SKA sensitivities can reach values of
$\mathcal{O}(10^{-30})\mathrm{cm^3 s^{-1}}$ within the mass range of $\mathcal{O}(1)-\mathcal{O}(10)$ GeV for the $e^{+}e^{-}$ channel. The sensitivities exhibit a higher magnitude in the mass range of  $\mathcal{O}(1)-\mathcal{O}(10)$ GeV for the $\mu^+\mu^-$ channel. For the $b\overline{b}$ channel, due to the b quark's mass at the GeV scale, the sensitivities derived from the synchrotron radiations manifest in the mass range above GeV. In the mass range from several GeV to TeV, the median sensitivities are about $\mathcal{O}(10^{-27})-\mathcal{O}(10^{-26})\mathrm{cm^3 s^{-1}}$. It is evident that the sensitivities for the $b\bar{b}$ channel are notably weaker than the leptonic channels for light DM. This disparity can be attributed to the exceedingly soft initial $e^\pm$ spectrum of the $b\bar{b}$ channel. As implied by Eq.~\ref{eq:ve}, $e^\pm$ particles with low energies would traverse considerable energy-loss distances, facilitating their fast escape from the dSph without a significant radiation rate, particularly given the limited diffusion region under consideration here. Consequently, for light DM, the $b\bar{b}$ channel yields substantially diminished radio signatures compared to the leptonic channels. From Fig.~\ref{Fig:sigma_v}, it is important to note the uncertainties associated with the DM density distribution, as the sensitivities may undergo fluctuation by a factor of  $\mathcal{O}(10)$ within the $68\%$ containment band. The fluctuation factor could exceed two orders of magnitude for certain extremely profiles within the $95\%$ containment band.

For the purpose of comparison, a variety of constraints derived from different observations have been included in the plot. The constraints obtained from the cosmic microwave background data~\cite{Cang:2020exa, Slatyer:2016qyl, Boudaud:2016mos}, the 15 dSphs observations of Fermi-LAT~\cite{Baring:2015sza, Fermi-LAT:2016uux}, and the cosmic-ray $e^\pm$ observations of Voyager and AMS02 ~\cite{Cang:2020exa, Slatyer:2016qyl, Boudaud:2016mos} for the $e^+e^-$ and $\mu^+\mu^-$ channels are displayed. 
The comparison demonstrates that the median sensitivities obtained in this study significantly surpass the constraints from previous studies by more than two orders of magnitude for DM masses below the 100 GeV. Even when accounting for the uncertainties related to the DM density profile, the sensitivities derived in this study remain notably superior to the constraints established by earlier research for the leptonic channels.
For the $b\bar{b}$ channel, the constraints derived from the 15 dSphs observations of Fermi-LAT~\cite{Baring:2015sza, Fermi-LAT:2016uux}, the combination of 20 dSphs observations performed by Fermi-LAT, HAWC, H.E.S.S., MAGIC, and VERITAS collaborations~\cite{Hess:2021cdp}, the 30 dSphs observations of Fermi-LAT~\cite{McDaniel:2023bju}, and the Ursa Major III observation of Fermi-LAT with a fixed J-factor of $J=10^{21} \rm GeV^2 cm^{-5}$~\cite{Crnogorcevic:2023ijs} are also shown for comparison.
The median sensitivities for the $b\bar{b}$ channel are superior to the majority of constraints from previous studies, with the exception of those derived from the Fermi-LAT observation of Ursa Major III with a fixed large J factor.

\begin{figure}[htbp]
\centering
\begin{subfigure}{0.48\textwidth}
    \centering
    \includegraphics[width=\textwidth]{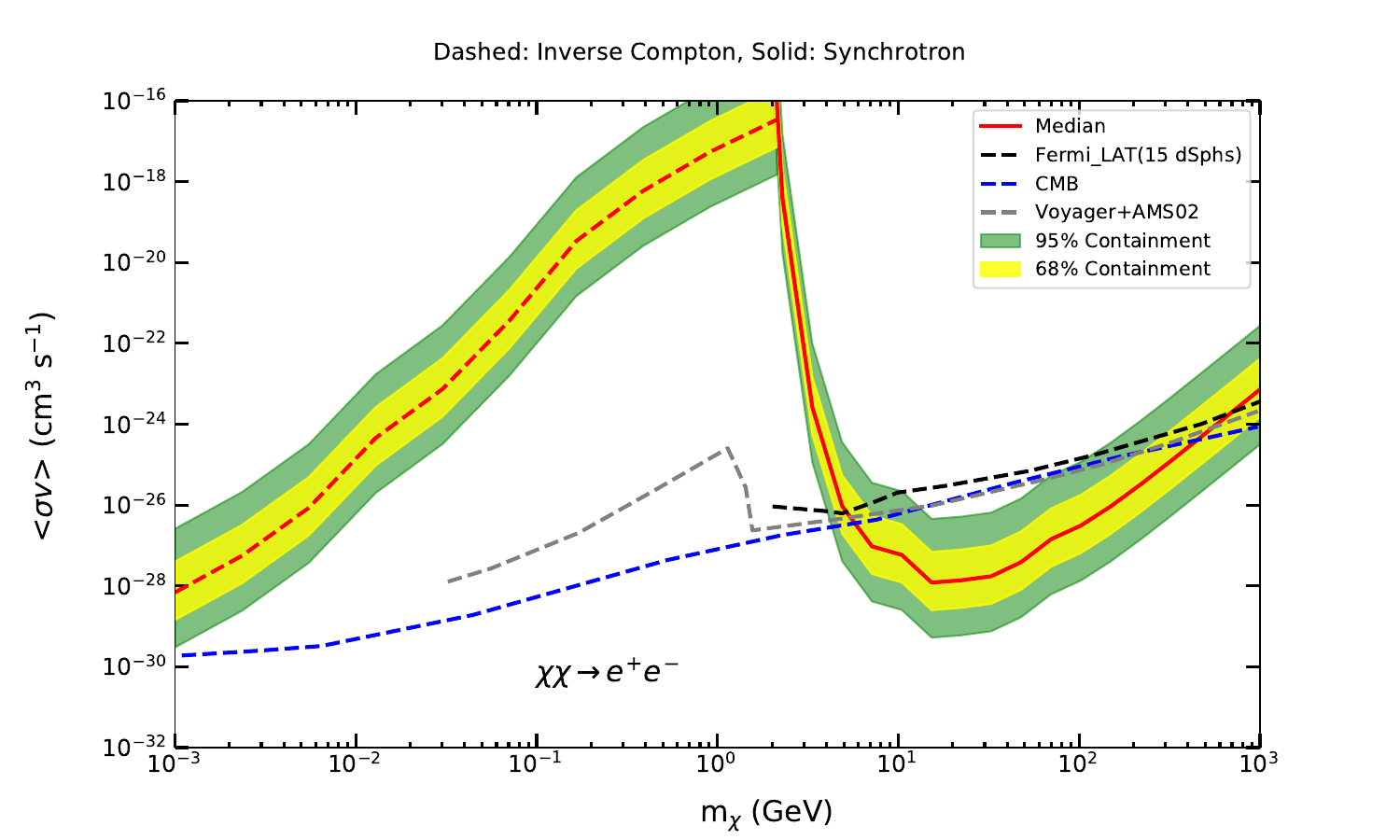}
    \subcaption{(a) $e^{+}e^{-}$ channel}
    \label{fig:ee_sigmav_subfig_a}
\end{subfigure}
\hfill
\begin{subfigure}{0.48\textwidth}
    \centering
    \includegraphics[width=\textwidth]{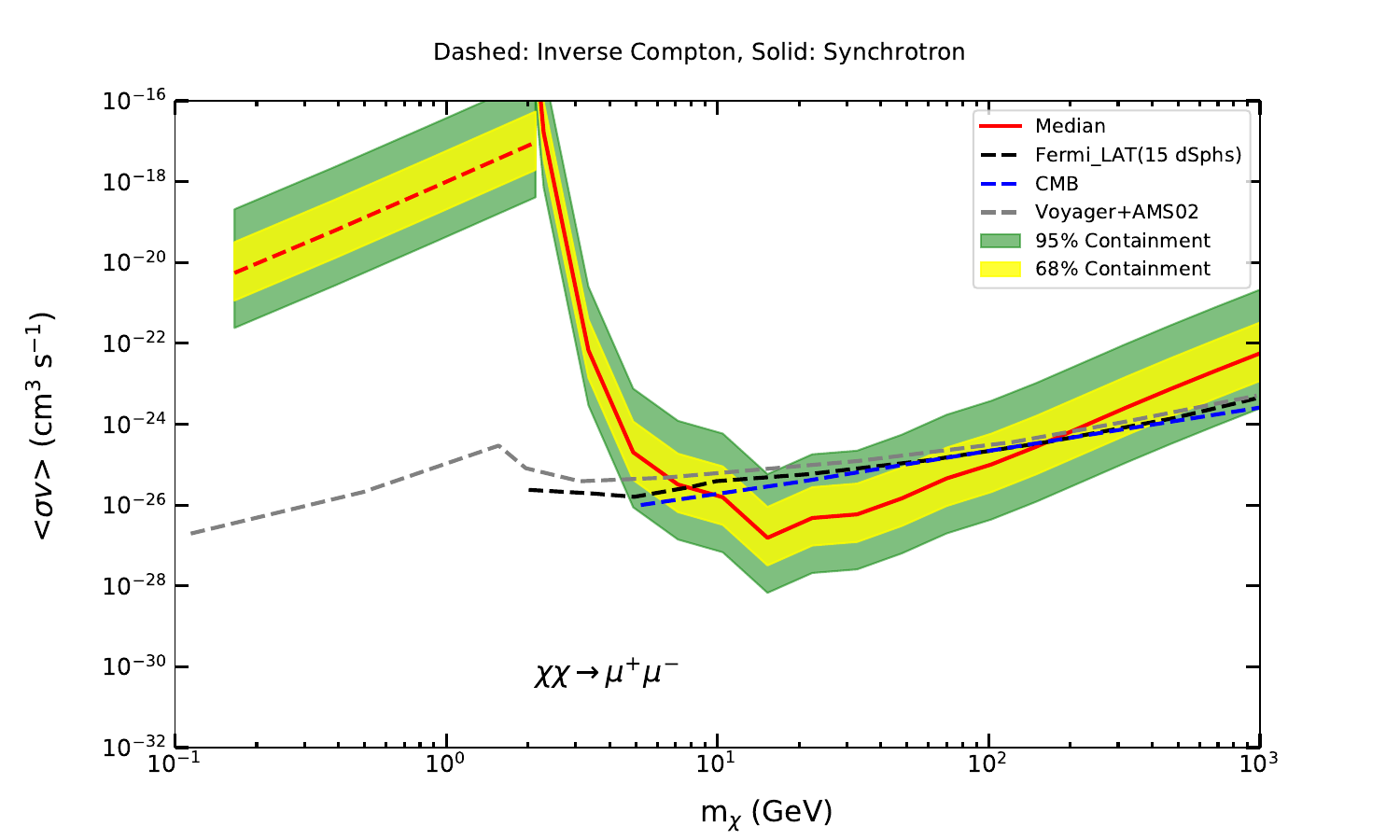}
    \subcaption{(b) $\mu^{+}\mu^{-}$ channel}
    \label{fig:mumu_sigamv_subfig_b}
\end{subfigure}
\hfill
\begin{subfigure}{0.48\textwidth}
    \centering
    \includegraphics[width=\textwidth]{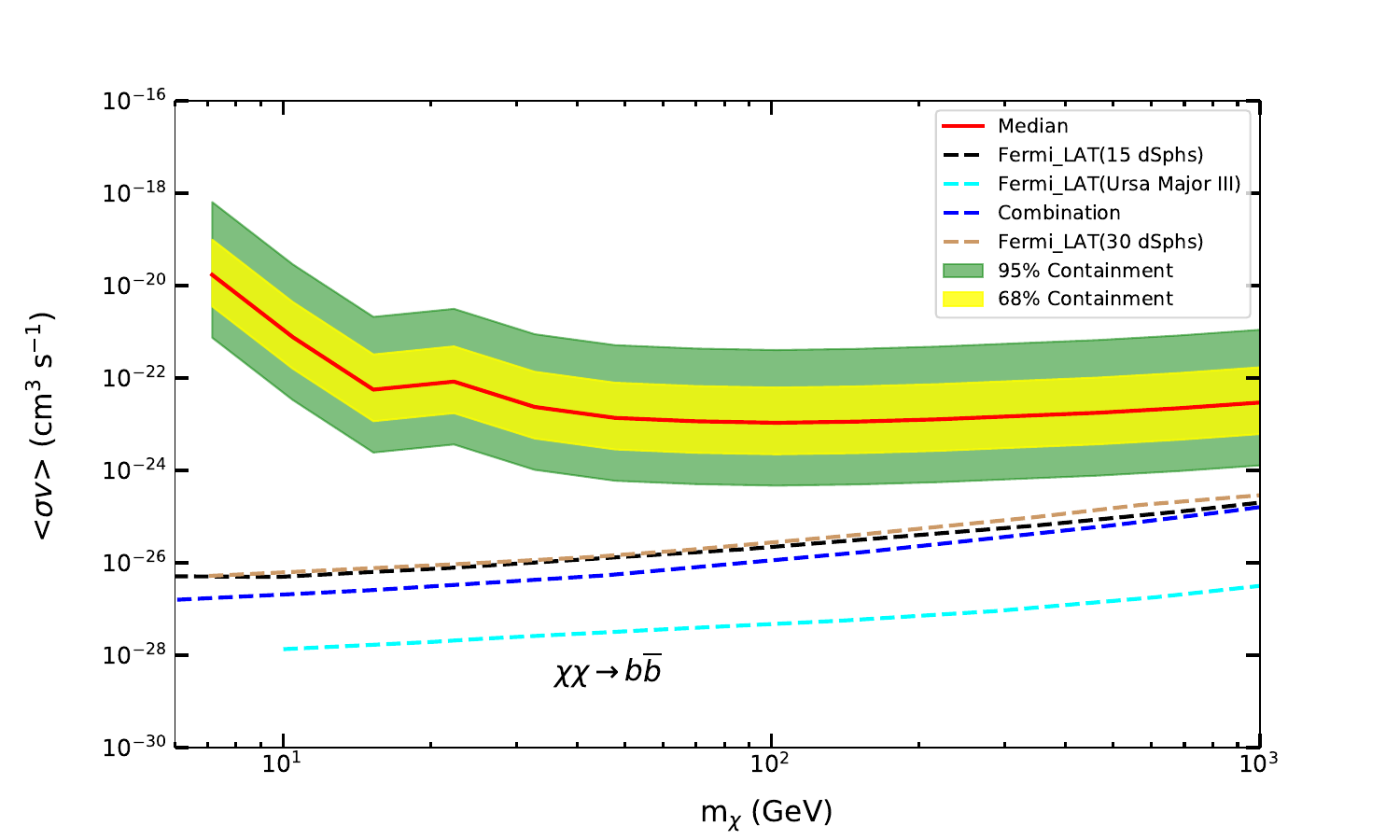}
    \subcaption{(c) $b\overline{b}$ channel }
    \label{fig:bb_bar_sigamv_subfig_c}
\end{subfigure}
\caption{Similar as Fig.~\ref{Fig:sigma_v_positive}, but for 
$B_{0}=0.1\mu$G and $D_{0} = 10^{30} \rm cm^2 s^{-1}$.
}
\label{Fig:sigma_v}
\end{figure}

The sensitivities of the SKA in the conservative scenario are illustrated in Fig.~\ref{Fig:sigma_v}. Due to the reduced magnetic field strength and increased diffusion coefficient in this scenario, the sensitivities are notably weaker compared to those in the optimistic scenario. Specifically, the sensitivities for the $b\bar{b}$ channel are above the existing constraints from other observations, suggesting that this channel may not be detectable in the conservative scenario. However, for the leptonic channels, the median sensitivities surpass the current constraints for DM masses from several GeV to 100 GeV. Even considering the uncertainty from the DM density profile, the sensitivities for the $e^+e^-$ channel still surpass the current constraints.
It is important to note that in the above analysis, we have fixed the radius of the diffusion region $r_h$ in Ursa Major III. Considering the potential substantial extension of the DM halo in this dSph, the self-confinement mechanism of $e^\pm$ resulting from DM annihilation may indicate a larger effective diffusion region compared to our current assumption. 
If the radius of the diffusion region $r_h$ is expanded, the sensitivities could be enhanced, as shown in  Fig.~\ref{Fig:SED_rh}.

\section{Conclusions}
\label{sec:conclusion}

The recent discovery of the Ursa Major III system has positioned it as potentially the faintest and densest dSph in the Milky Way. With a substantial DM component and its proximity to Earth, Ursa Major III presents a very promising target for DM indirect searches. In this study, we investigate the sensitivities of the SKA to the radio signatures arising from DM annihilation within Ursa Major III  for the 
$e^{+}e^{-}$, $\mu^{+}\mu^{-}$, and  $b\bar{b}$ annihilation channels.
We calculate the SEDs resulting from the synchrotron and ICS processes, and examine the influences of uncertainties in astrophysical parameters. Notably, an increase in the magnetic field strength and a reduction in the diffusion coefficient could substantially amplify the radio flux originating from synchrotron radiation. Furthermore, an increase in the radius of the diffusion region would lead to a significant enhancement in the radio flux.

In our analysis, we calculate the median sensitivities of the SKA, along with their 68\% and 95\% containment bands, accounting for uncertainties of the DM density profile in Ursa Major III. Two distinct scenarios are are considered: an optimistic scenario with $B_0 = 1\mu\rm G$ and $ D_0= 3\times 10^{28} \mathrm{cm^2 s^{-1}}$, and a conservative scenario with $B_0 = 0.1\mu\rm G$ and $ D_0=  10^{30} \mathrm{cm^2 s^{-1}}$. Given the substantial DM component and proximity of Ursa Major III, the SKA exhibits robust capabilities in detecting synchrotron radiations originating from DM annihilation in the optimistic scenario. For instance, the sensitivities to the DM annihilation cross section can reach values of
$\mathcal{O}(10^{-30})\mathrm{cm^3 s^{-1}}$ in the mass range of $\mathcal{O}(1)-\mathcal{O}(10)$ GeV for the $e^{+}e^{-}$ channel. These sensitives surpass the current most stringent limits derived from cosmic microwave background, gamma-ray, and cosmic-ray $e^\pm$ observations. Even when considering uncertainties in the DM density profile of Ursa Major III, the SKA sensitivities remain significant, albeit experiencing a reduction for certain density profiles. In the conservative scenario, where the magnetic field strength is lower and the diffusion coefficient is higher, the SKA sensitivities are weakened. Nevertheless, the SKA still demonstrates significant capabilities to investigate light DM for the  leptonic channels.

It is imperative to note that the precise DM content in Ursa Major III remains undetermined. Our analysis is based on current kinematic observations of 11 member stars within Ursa Major III. Variations in kinematic data could lead to a significant reduction in the inferred DM content, potentially diminishing the sensitivities derived in this study. Therefore, further elucidation of kinematic data is crucial for a comprehensive investigation into DM signatures originating from Ursa Major III.

%\newpage

\section*{Acknowledgments}
We would like to thank Zhan-Fang Chen, Guan-Sen Wang, and En-Sheng Chen for valuable discussions. 
This work is supported by the National Natural Science Foundation of China (Grant No. 12175248, 12275067), Natural Science Foundation of Henan Province (Grant No. 225200810030), and Excellent Youth Foundation of Henan Province (Grant No. 2123004140010).

\bibliography{SKA_DM}

\end{document}